\documentstyle[11pt,epsfig,array,epsf]{article}

\begin{document}
\renewcommand\topfraction{0.1}
{\normalsize \renewcommand\baselinestretch{1.1}
\title{Elastic properties of a tungsten-silver composite by
reconstruction and
computation\footnote{Submitted to J.\ Mech.\ Phys.\ Solids}}
\author{A. P. Roberts\footnote{Permanent address: Centre for
microscopy and microanalysis, University of Queensland, Brisbane,
Queensland 4072, Australia}
\\
University of Oxford, Department of Materials, \\
Parks Rd, Oxford OX1--3PH, U.K. \\
and \\
E. J. Garboczi \\
Building Materials Division, \\
National Institute of Standards and
Technology, \\ 100 Bureau Drive Stop 8621, Gaithersburg, MD 20899-8621, U.S.A.}
\date{\today}
\maketitle
}


\begin{abstract}
We statistically reconstruct a three-dimensional model
of a tungsten-silver composite from an experimental two-dimensional
image. 
The effective Young's modulus ($E$) of the model is computed in
the temperature range 25--1060$^o$C using a finite element method.
The results are in good agreement with experimental data.
As a test case, we have reconstructed the microstructure and 
computed the moduli of the 
overlapping sphere model.
The reconstructed and overlapping sphere models are examples of bi-continuous
(non-particulate) media.
The computed moduli of the models are not generally in good
agreement with the predictions of the self-consistent method.
We have also evaluated three-point variational bounds on
the Young's moduli of the models using the results of Beran, Molyneux,
Milton and Phan--Thien. The measured data were close to the
upper bound if the properties of the two phases were similar ($\frac{1}{6} < E_1/E_2 < 6$).
\end{abstract}

\newcommand{\iny}{$\infty$}
\section{Introduction}

Predicting the macroscopic properties of
composite or porous materials with random
microstructures is an important problem in
a range of fields~\cite{Bergman78,Hashin83,TorqRev91,Sahimi93}.
There now exist large-scale 
computational methods for calculating the properties
of composites  given a
digital representation of their microstructure;
eg.\ permeability~\cite{Bentz94a,Adler90},
conductivity~\cite{Adler92,Roberts95a}
and elastic moduli~\cite{Garboczi95a,Adlerelas1}. 
A critical problem is obtaining an accurate
three-dimensional (3D) description of this
microstructure~\cite{Bentz94a,Crossley91,Yao93}.

For particular materials it may be possible to simulate
microstructure formation from first principles.
Generally this relies on detailed knowledge of the
physics and chemistry of the system, with accurate
modeling of each material requiring a significant amount
of research. Three-dimensional models have also
been directly reconstructed from samples by combining
digitized serial sections obtained by scanning
electron microscopy \cite{Kwiecien90},
or using the relatively new technique of x-ray
microtomography~\cite{Flannery3DXM}.
In the absence of sophisticated experimental facilities,
or a sufficiently
detailed description of the microstructure formation (for
computer simulation), a third alternative is
to employ a statistical model of the microstructure.
This procedure has been termed ``statistical reconstruction''
since the statistical properties of the model are
matched to those of a
two-dimensional (2D) image~\cite{Quiblier84,Adler92,Bentz94a,RobertsRec}.
Statistical reconstruction is a promising
method of producing 3D models, but there remain outstanding
theoretical questions regarding its application. First, what is the
most appropriate statistical information (in a 2D image) for reconstructing
a 3D image, and second, is this information sufficient to
produce a useful model? In this paper we address these
questions, and test the method against experimental data.

Modeling a composite and numerically estimating its
macroscopic properties is a complex procedure. This could
be avoided if accurate analytical structure-property
relations could be theoretically or empirically obtained.
Many studies have focussed on this problem~\cite{Hashin83}.
In general, the results are reasonable for a particular class
of composites or porous media.
The self-consistent (or effective medium) method of
Hill~\cite{HillSCM} and Budiansky~\cite{BudianskySCM}
and its generalization by Christensen and Lo~\cite{ChristensenLoGSCM}
is one of the most common
for particulate media~\cite{Hashin83}. No analogous results are available
for non-particulate composites.
A promising alternative to direct property prediction has
been the development of analytical rigorous
bounds (reviewed by Willis~\cite{Willis81}, Hashin~\cite{Hashin83}
and Torquato~\cite{TorqRev91}).
There is a whole hierarchy of these bounds, each set tighter
than the next, but depending on higher and higher order
correlation functions of the microstructure.
The original Hashin and Shtrikman~\cite{HSelas}
bounds that have been widely used by experimentalists
implicitly depend on the two-point correlation function of
the microstructure, although the only quantities appearing
in the formulas are the individual properties of each phase and
their volume fractions.
To go beyond these bounds to higher-order, more restrictive (i.e., narrower)
bounds, it is necessary that detailed information be known about the composite
in the form of three-point or higher statistical correlation
functions~\cite{Beran65b,Milton82b}, which do appear explicitly 
in the relevant formulas.
Evaluation of even the three point function is a formidable task, so
use of these bounds has in the past been restricted to composites with
spherical inclusions. It is now possible to evaluate the bounds for
non-particulate composites~\cite{Roberts95a}, and it is interesting
to compare the results with experimental and numerical data.
If the properties of each phase are not too dissimilar the
bounds are quite restrictive and can be used for predictive
purposes~\cite{HSelas}.  Sometimes
experimental properties closely follow one or the other of the bounds,
so that the upper or lower bound often provides a reasonable
prediction of the actual property even when the phases have
very different properties~\cite{TorqRev91,Roberts95b}.
It is useful to test this observation.

In this study we test a generalized version~\cite{RobertsRec}
of Quiblier's~\cite{Quiblier84}
statistical reconstruction procedure
on a well-characterized silver-tungsten composite.
Computational estimates of the Young's moduli
are compared to experimental measurements. The composite is 
bi-continuous (both phases are macroscopically connected) and therefore
has a non-particulate character. As such the microstructure is
broadly representative of that observed in open-cell foams (such as
aerogels), polymer blends, porous rocks, and cement-based materials. 
By comparing our computations of the moduli to the results
of the self-consistent method we can test its utility for
non-particulate media.
An advantage of the reconstruction procedure we use is that it
provides the statistical correlation functions necessary for
evaluating the three-point bounds.
Comparison of the Young's modulus to the bounds therefore allows
us to determine the bounds' range of application for
predictive purposes.

\section{Statistical models of microstructure}
\label{statmod}

The two basic models we employ to describe two-phase composite microstructure
are the overlapping sphere model and the level-cut
Gaussian random field (GRF) model.
In this section we review the statistical
properties of these models which are useful for reconstructing
composites.
The simplest, and most common, quantities used to characterize
random microstructure are $p$, the volume fraction
of phase 1, $s_v$, the surface area to total volume ratio and
$p^{(2)}(r)$, the two-point correlation function
(or $\gamma(r)\equiv[p^{(2)}(r)-p^2]/[p-p^2]$ the auto-correlation
function).
$p^{(2)}(r)$ represents the probability that two points
a distance $r$ apart lie in phase 1.
Here we only consider isotropic materials where $p^{(2)}$ does
not depend on direction.
Also note that $p=p^{(2)}(0)$ and $s_v=-4dp^{(2)}(0)/dr$.

Realizations of the overlapping
sphere model~\cite{Weissberg63} are generated by
randomly placing spheres (of radii $r_0$) into a matrix.
The correlation function of the phase exterior
to the spheres (fraction $p$) is
$p^{(2)}(r)=p^{v(r)}$ for $r<2r_0$ and
$p^{(2)}(r)=p^2$ for $r\geq 2r_0$ where
\begin{equation}
v(r)=1+\frac34 \left(\frac r{r_0}\right)
      -\frac1{16}\left(\frac r{r_0}\right)^3,
\label{p2ios}
\end{equation}
and the surface to volume ratio is $s_v=-3p\ln p/r_0$. 
With modification it is also possible to incorporate
poly-dispersed and/or hollow spheres.
The overlapping sphere model is the most well characterized
of a wider class called Boolean models, which have been recently
reviewed by Stoyan {\em et al.}~\cite{StoyanSG}.

The internal interfaces of a different class of
composites can be modeled by the iso-surfaces
(or level-cuts) of a stationary correlated
Gaussian random field (GRF) $y(\mbox{\boldmath$r$})$
(so called because the value of the field at
randomly chosen points in space is Gaussian--
distributed). Moreover, if $\mbox{\boldmath$r$}$ is fixed, the
distribution over an ensemble will also be Gaussian.
Correlations in the field are governed by the
field-field correlation function
$g(r)=\langle y(0)y(\mbox{\boldmath$r$})\rangle$ which
can be specified subject to certain constraints
[$|g(r)|<g(0)$, $\lim_{r\to\infty}g(r)\to0$]. Invariably
$g(0)$ is taken as unity. A useful general
form for $g$ is~\cite{RobertsRec}
\begin{equation}
g(r)=\frac{e^{-r/\xi}-(r_c/\xi)e^{-r/r_c}}{1-(r_c/\xi)}
\frac{\sin 2\pi r /d}{2\pi r /d}.
\label{modelg}
\end{equation}
The resulting field is
characterized by a correlation length $\xi$, domain scale
$d$ and a cut-off scale $r_c$.
The cut-off scale is necessary to
ensure $1-g(r) \sim r^2$ as $r\to0$; fractal iso-surfaces are generated
if $1-g(r) \sim r$.
There are many algorithmic methods of generating random
fields. A straight forward method is to sum $N$ ($\sim$1000)
sinusoids with random phase and wave-vector
\begin{equation}\label{ydefn}
y(\mbox{\boldmath $r$})=\sqrt{\frac{2}{N}}\sum_{i=1}^{N}
\cos(k_i \hat{\mbox{\boldmath $k$}}_i \cdot {\mbox{\boldmath $r$}} + \phi_i),
\end{equation}
where $\phi_i$ is a uniform deviate on $[0,2\pi)$ and
$\hat{\mbox{\boldmath $k$}}_i$ is uniformly distributed on a unit sphere.
The magnitude of the wave vectors $k_i$ are distributed on
$[0,\infty)$ with a probability (spectral) density $P(k)$. The
density is related to $g(r)$ by a Fourier transform
[$g(r)$=$\int_0^\infty P(k)\sin kr (kr)^{-1} dk$]. Note that $P(k)>0$
specifies an additional constraint on $g(r)$. Although this formulation
of a GRF is intuitive, the Fast Fourier Transform
method is more efficient~\cite{Adlerbook,Roberts95a}.

Following Berk \cite{Berk87} one can define a composite with
phase 1 occupying the region in space where
$\alpha\leq y(\mbox{\boldmath$r$}) \leq \beta$ and phase 2 occupying the
remainder.
The statistics
of the material are completely determined by the specification of
the level-cut parameters and the function $g(r)$ (or $P(k)$).
The volume fraction of phase 1 is 
\begin{equation}
p=h=p_\beta-p_\alpha\;\;\;\;{\rm where}\;\;\;\;
p_\gamma=(2\pi)^{-\frac12}\int_{-\infty}^\gamma e^{-t^2/2} dt\;, \;\;\gamma = \alpha, \beta.
\label{gamma}
\end{equation}
Berk~\cite{Berk87} and Teubner~\cite{Teubner91} have shown that the
two point correlation function is $p^{(2)}(r)=h(r)$
where \begin{eqnarray}
h(r)= && h^2+\frac{1}{2\pi}\int_0^{g(r)}
 \frac{dt}{\sqrt{1-t^2}} \times  \left[
\exp\left({-\frac{\alpha^2}{1+t}}\right) \right.
\label{h2}
\\ && \left. \nonumber
-2\exp\left({-\frac{\alpha^2-2\alpha\beta
t+\beta^2}{2(1-t^2)}}\right)
+\exp\left({-\frac{\beta^2}{1+t}}\right) \right].
\end{eqnarray}
The auxiliary variables $h$ and $h(r)$ are needed below. 
The singularity at $t=1$ can be removed with the substitution $t=\sin\theta$.
The specific surface is $s_v=-4h'(0)$ where
\begin{equation}
-h'(0) = \frac{\sqrt{2}}{2\pi}\left( e^{-\frac12 {\alpha^2}}+
e^{-\frac12 \beta^2} \right)\sqrt{\frac{4\pi^2}{6d^2}+\frac{1}{2r_c\xi}}
\label{hdash}
\end{equation}
with $g$ given by Eqn.~(\ref{modelg}).

Many more models (for which $p^{(2)}(r)$ can be simply evaluated)
can be formed from the intersection and union sets of the overlapping sphere
and level-cut GRF models. Here we define a few representative models
which have been shown to be applicable to composite and
porous media. 
A normal model (N) corresponds to Berk's formulation.
Models can also be formed from the
intersection (I) and union (U) of two statistically identical
level-cut GRF's.  Another model, I$_n$, formed
from the intersection of $n$ primary models, has also been found
useful.
The statistical properties ($p$, $s_v$ and $p^{(2)}$) of each model
are given in terms of the properties of Berk's model [Eqns.~(\ref{gamma}),
(\ref{h2}) and (\ref{hdash})] in 
Table~\ref{tabmods}~\cite{RobertsRec}.

\begin{table}[bt!]
\caption{
The volume fraction $p$, two-point correlation function
$p_2(r)$ and surface
to volume ratio $s_v$ of models N, I, U and I$_n$
in terms of the properties [$h$, $h(r)$ and $h'(0)$]
of Berk's two-level cut Gaussian random field model.
The formula $h_p$ is used for calculating the level-cut parameters
(see Table~\protect\ref{tabab}).  \label{tabmods}}
\begin{center}
\setlength{\extrarowheight}{2pt}
\begin{tabular}{|c|c|c|c|c|}       
\hline
Mod.\ & $p$ & $p^{(2)}(r)$ & $s_v$ & $h_p$
\\
\hline
N    & $h$            & $h(r) $           & $-4h'(0)$ & $p$ \\
I    & $h^2$     & $h^2(r)$          & $-8hh'(0)$ & $\sqrt{p}$ \\
U    & $h(2-h)$ & $[2h^2+2h(r)$      & $-8(1-h)h'(0)$ & $1-\sqrt{1-p}$ \\
     &                & $-4hh(r)+h^2(r)]$ &                &  \\
I$_n$& $h^n$      & $h^n(r)$          & $-4nh^{n-1}h'(0)$ & $p^{1/n}$\\
\hline
\end{tabular}
\end{center}
\end{table}

Since the volume fraction of the models is a function of
both level-cut parameters ($\alpha$,$\beta$) there is a continuum of
choices which correspond to a given volume fraction. For example
$(\alpha,\beta)$=$(-\infty,0.84)$, $(-0.84,-0.25)$,
$(-0.25,0.25)$, $(0.25,0.84)$ and $(0.84,\infty)$ in Berk's model
all correspond to $p=20\%$. The final two choices are statistically identical
to the first two and therefore provide nothing new. We note that a small
change in these parameters will only slightly alter the microstructure,
so as a compromise between simplicity and generality it is
suggested only three distinct cases be considered: (i) the common single-cut
field ($\alpha=-\infty$); (ii) a symmetric two-cut field ($\alpha=-\beta$) and
(iii) an asymmetric two-cut field.
A concise way of expressing this is to
take $p_\alpha=\frac c2 - \frac c2 h_p$ and $p_\beta=\frac c2 + (1-\frac c2)
h_p$ where $h_p$ for models N, I, U and I$_n$ is given in
Table~\ref{tabmods} and $c\in[0,1]$.
Setting $c$=0, 1 and $\frac12$ gives cases (i), (ii) and (iii) respectively. 
The implicit formula for finding $(\alpha,\beta)$ from $p_\alpha$ and 
$p_\beta$ is shown in Eqn.~(\ref{gamma}). As an example
the results of the calculation for nine different models
(N, I and U at $c$=0, $\frac12$ and 1) at volume fraction 20\%
is shown in  Table~\ref{tabab}. The model N ($c$=0) is the single
level cut GRF previously used by Quiblier~\cite{Quiblier84} and
Teubner~\cite{Teubner91}.
Model I ($c$=1) has been used to model aerogels~\cite{RobertsAero}
and model N ($c$=1) is Berk's~\cite{Berk87} two level
cut model of microemulsions.

\begin{table}[bt!]
\caption{The level-cut parameters $\alpha$ and $\beta$ for different Gaussian
random field models are calculated by solving Eqn.~(\protect\ref{gamma}) where
$p_\alpha=\frac c2 - \frac c2 h_p$ and $p_\beta=\frac c2 + (1-\frac c2)
h_p$. $h_p$ is shown in Table~\protect\ref{tabmods} for each model
and $c=0,\frac12,1$.  This table shows the results of the calculation for
volume fraction $p$=20\%. 
\label{tabab}}
\begin{center}
\begin{tabular}{|c|cc|cc|cc|}       
\hline
\multicolumn{1}{|c|}{Model} &
\multicolumn{2}{c|}{Standard} &
\multicolumn{2}{c|}{Asymmetric} &
\multicolumn{2}{c|}{Symmetric} \\
\multicolumn{1}{|c|}{type} &
\multicolumn{2}{c|}{one-cut} &
\multicolumn{2}{c|}{two-cut} &
\multicolumn{2}{c|}{two-cut} \\
\hline
\multicolumn{1}{|c|}{} &
\multicolumn{2}{c|}{$c=0$} &
\multicolumn{2}{c|}{$c=1/2$} &
\multicolumn{2}{c|}{$c=1$} \\
\hline
     & $\alpha$ & $\beta$ & $\alpha$ & $\beta$  & $\alpha$ & $\beta$ \\
\hline
Normal (N)      & -$\infty$ & -0.84 & -0.84 & -0.25&-0.25&  0.25 \\
Intersection (I)& -$\infty$ & -0.13 & -1.09 &  0.22&-0.59&  0.59 \\
Union (U)       & -$\infty$ & -1.25 & -0.76 & -0.44&-0.13&  0.13 \\
\hline
\end{tabular}
\end{center}
\end{table}

\section{Statistical reconstruction}
\subsection{The Joshi-Quiblier-Adler (JQA) Approach}
\label{statrec}

The basic idea of statistical reconstruction is to generate
a three dimensional (3D) model of a random composite using only
statistical information measured from a two
dimensional (2D) image. Since 2D cross-sections are readily
obtained by common experimental methods this provides
a very attractive method of modeling porous and composite media.
Quiblier~\cite{Quiblier84} developed a method
capable of producing a three-dimensional model with
a specified volume fraction and two-point correlation function.
The method was first studied in two dimensions by Joshi~\cite{JoshiThesis}
and has been extended by Adler~\cite{Adlerbook}.  Thus it
has been called the JQA model.
We first give a summary of the procedure to
demonstrate its equivalence to the single-cut GRF model
discussed in the last section. First $p$ and $p^{(2)}_{\rm expt}$
are measured from an experimental image. The level-cut parameter $\beta$ is
fixed by the volume fraction [Eqn.~(\ref{gamma}) with $\alpha=-\infty$]. 
Second $g_{\rm expt}(r_i)$ is obtained at the set of
discrete points where $p^{(2)}_{\rm expt}$ is measured by inverting
Eqn.~(\ref{h2}) with
the left hand side set to $p^{(2)}_{\rm expt}(r_i)$ and $\alpha=-\infty$.
To carry out the inversion the right hand side of Eqn.~(\ref{h2}) is expanded
as a series in powers of $g$;
\begin{equation}
p^{(2)}_{\rm expt}(r_i)=p^2+p(1-p)\sum_{m=0}^\infty C_m^2 [g_{\rm expt}(r_i)]^m
\end{equation}
where the coefficients $C_m$ depend on certain integrals of Hermite
polynomials~\cite{Quiblier84}.
The series converges very slowly for $g$ near one, and
Alder~\cite{Adlerbook} has discussed the inversion procedure in
this case, and possible situations where the equation has
no solution.
We note that the inversion can also be simply carried out
by numerical integration of Eqn.~(\ref{h2}) and a standard
non-linear equation solver.

Next it is
necessary to generate a GRF with field-field correlation
function $g_{\rm expt}(r)$. Quibler's original formulation involved
the solution of a very large system of non-linear equations for
the terms of a convolution operator used in his
definition of a GRF. Adler~\cite{Adlerbook}
simplified the procedure by reformulating the problem in
terms of Fourier transforms, which is equivalent
to a three-dimensional version~\cite{Roberts95a} of
Rice's~\cite{RiceI} method for Gaussian processes in one dimension.
In terms of the definition given in Eqn.~(\ref{ydefn})
the inversion is equivalent to a numerical integration of
$P(k)=\frac{2}{\pi}\int_0^\infty g(r) rk \sin kr dr$,
where $g$ is known only at a discrete set of points.
The inversion methods described above for $g(r)$ do not
guarantee that $P(k)$ (or its equivalent in other formulations)
is greater than zero.  However Adler~\cite{Adlerbook}
found that if $P(k)$ is negative at some points it is also small,
and can be replaced with zero. Finally the GRF
is thresholded in the usual way to obtain the reconstructed
microstructure. Thus the JQA method produces a single-cut
Gaussian random field, which we have termed model N($c$=0).

In this paper we employ a different implementation of
the JQA method~\cite{RobertsRec}. At this stage we restrict
attention to model N($c$=0).
First, the volume fraction of the model is set to
that of an image: $p_{\rm{mod}}$=$p_{\rm{expt}}$.
Second, the experimental two point correlation function is
fitted by varying the morphological parameters of a {\em given}
$g(r)$ [Eqn.~(\ref{modelg})] to minimize the non-linear least squares error
\begin{equation}
[Ep^{(2)}]^2 = 
\frac{\sum_{i=1}^{N_f} [p^{(2)}_{\rm{mod}}(r_i)-p^{(2)}_{\rm{expt}}(r_i)]^2}
{\sum_{i=1}^{N_f} [p^{(2)}_{\rm{expt}}(r_i)-p_{\rm{expt}}^2]^2}.
\label{Ep2}
\end{equation}
where $N_f$ is the number of experimental points to be fitted.
Numerical integration is used to
find $p^{(2)}_{\rm{mod}}(r_i)$ [Eqn.~(\ref{h2})].
The minimization is very fast, but several starting points 
should be used as a check against local minima.
Once the parameters of $g(r)$ are known an analytic form of
$P(k)$ is used to generate the coefficients of the GRF (\ref{ydefn}).
The reconstructed model is obtained by thresholding the
random field as described above. Note that
$p^{(2)}_{\rm{mod}}(r_i)$ will not match $p^{(2)}_{\rm{expt}}(r_i)$
exactly at each point as would be the case with Quiblier's procedure.
However with this choice of $g(r)$, $P(k)$ is guaranteed to be
positive. More general functional forms of $g(r)$ can also be employed.

\begin{figure}[t!]
\centering\epsfig{figure=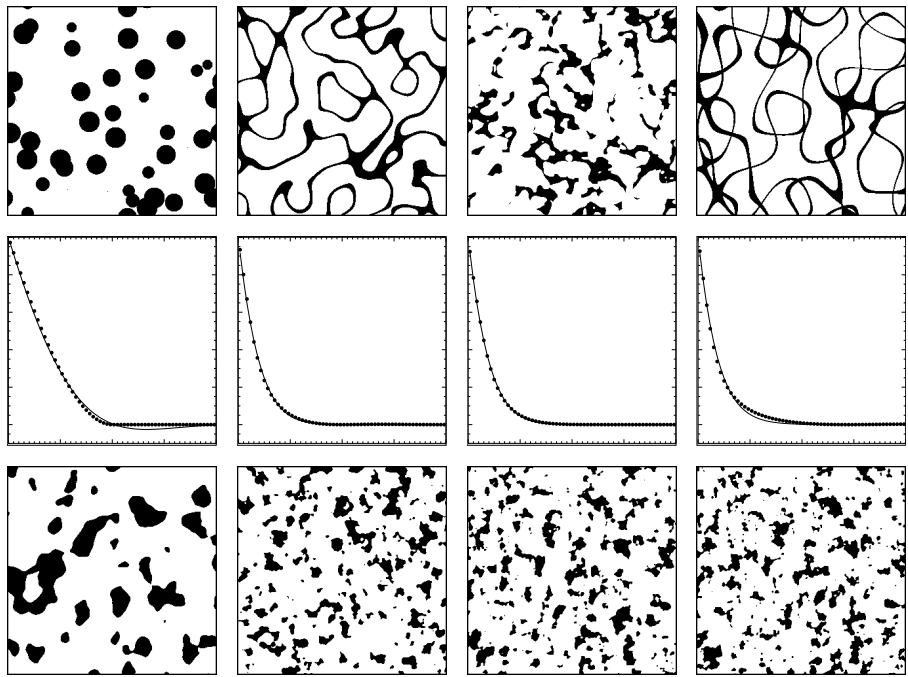,width=12.0cm}
\caption{The statistical reconstructions (bottom row)
of four different two-dimensional images by
adjusting the parameters ($r_c$, $\xi$, $d$ and $\beta$)
of model N($c$=0) to fit the auto-correlation functions
(middle row) of the original models (shown in the top row).
The procedure used is very similar to that of Quiblier.
The results suggest that model N($c$=0) cannot mimic all types
of microstructure, even though it can reproduce the
two-point correlation functions of all the cases shown.
The models in the top row are (from left to right), overlapping spheres,
models N($c$=1), I($c$=1) and U($c$=1) (see Sec.~\protect\ref{statmod}). 
The images are 128$\times$128 pixels, and the length scale of the
correlation functions shown in the 2nd row extends to 32 pixels.
\label{all_same}} 
\end{figure}

For isotropic materials $p$ and  $p^{(2)}(r)$ can be
exactly measured from a two (or one) dimensional image.
Hence the application of the JQA method results in
a model which shares $p$, $s_v$ and $p^{(2)}(r)$ with
a real composite. The question is whether or not
the model provides an accurate and useful representation
of the original microstructure. In certain cases it appears to,
in other it does not.
First, predictions of transport properties (conductivity and
permeability) obtained from reconstructed porous models
under-estimate experimental and numerical data. 
Second, the percolation threshold (the volume fraction at which
the pore space or inclusion phase is no longer macroscopically
connected) of model N($c$=0) is around 10\%~\cite{Roberts95a}
for the model, but many materials exhibit lower thresholds~\cite{Roberts96a}.
Both points indicate that the pores (or inclusions) of the model are not
sufficiently well connected to mimic many physical materials.
Third, we can test the model by trying to reconstruct several of the
distinct models defined in the previous section. The results are
shown in Fig~\ref{all_same}. Even though model N($c$=0)
is able to reproduce the correlation functions reasonably well,
the reconstructions (i) do not in all cases appear to reproduce
the original microstructure and (ii) look
quite similar to one another. This indicates that irrespective of
$g(r)$, and the original image, model N($c$=0) can
only generate microstructures that are similar to those shown in
the final row of Fig.~\ref{all_same}.

Quiblier~\cite{Quiblier84} has suggested
that the ability of the method to generate a reasonable model
depends on the validity of the hypothesis that:
{\em all the necessary information about the morphology
is contained in the auto-correlation function}. 
Suppose this were true. Then since the JQA method is sufficiently
general to re-produce all reasonable two-point correlation
functions (see Fig.~\ref{all_same}), it must also be
able to generate all types of morphology.
The discussion above (and Fig.~\ref{all_same}) indicates that model N($c$=0)
can only produce a limited class of microstructure
and therefore that the hypothesis is false.
This does not mean that the the method cannot produce useful
models, but we argue it will do so only when the original
material is approximately contained in the same
limited class. If it is not, then
a model from a different class needs to be considered.
We discuss this issue in the following section.

\subsection{Using several models}

From the foregoing discussion it is clear that a model more general
than a single level-cut random field is needed to reproduce
the random isotropic microstructures seen in many
composite materials. At present there is no one model that can
achieve this. Instead it has been proposed that a number of
morphologically distinct models (N, I and U of Sec.~\ref{statmod})
be incorporated~\cite{RobertsRec}.
This is very simple to do by using the relevant formula
for $p^{(2)}_{\rm mod}(r)$ (Table~\ref{tabmods}) in Eqn.~(\ref{Ep2}).
It was found that many of these models were able to
match any given $p^{(2)}_{\rm expt}(r)$ (providing further evidence that
the two-point correlation function does not provide sufficient
information for a useful reconstruction). The problem
then becomes how to choose the best model. Clearly higher-order
statistical properties need to be taken into account.

The quantities $p$ and $p^{(2)}(r)$ are the first and second of
an infinite hierarchy of correlation functions,
the three-point function $p^{(3)}(r,s,t)$
representing the probability that three points, distances $r$, $s$ and $t$
apart fall in phase 1, and so forth.
Two random composites can only be said to be statistically identical
if their $N$-th order correlation functions
($N=1,2,3,4\dots$) are identical~\cite{Yao93}.
Therefore the exact statistical reconstruction of
a composite requires matching all of the correlation
functions\footnote{Technically, the model and composite may still
differ by a point-process~\cite{StoyanSG}, but this is unlikely to effect the
macroscopic properties.}.
An obvious method of choosing the best of several
models is then to compare $p^{(3)}$ of each model to experimental
data. As well as being memory and time intensive~\cite{Berryman_p2p3}
it has been shown~\cite{RobertsRec}
that $p^{(3)}$, like $p^{(2)}$, may not
contain the relevant morphological information (ie.\ it does not
provide a strong signature of microstructure).
A comparative study of several high-order statistical
quantities found that the simplest and most discriminating
signature of microstructure was the chord-distribution
functions for
each phase $\rho^{(j)}(r)$ ($j=1,2$). $\rho^{(j)}(r)$
is the probability that a randomly chosen chord in phase
$j$ has length $r$. A chord is defined as any line-segment which lies
entirely in phase $j$ with end points at the phase interface.
Like $p^{(2)}$ the chord functions are the same whether measured
from a two or three dimensional element of the microstructure.

The chord-functions can be employed in a reconstruction algorithm
as follows. First the morphological parameters [the length scales
in Eqns.~(\ref{p2ios}) and~(\ref{modelg})] of each model (overlapping spheres,
or the the various level-cut GRF's) are chosen to fit $p^{(2)}_{\rm expt}$.
We have found that most models are able to provide a
reasonable fit of $p^{(2)}_{\rm{expt}}(r)$ (e.g.\ $Ep^{(2)}<0.1$).
If this is not the case the model is unlikely to provide
a useful reconstruction and may be rejected.
Second, of the remaining candidates, the model that
best reproduces the experimental chord functions $\rho^{(j)}_{\rm expt}$
is selected as the best reconstruction.
We quantify the error by a normalised
least square sum;
\begin{equation}
\label{Epcd}
[E\rho^{(j)}]^2 = \frac{\sum_{i=1}^{M}
[\rho^{(j)}_{\rm rec}(r_i)-\rho^{(j)}_{\rm expt}(r_i)]^2}
{\sum_{i=1}^{M} [\rho^{(j)}_{\rm expt}(r_i)]^2}
\end{equation}
where $\rho^{(j)}_{\rm{rec}}$ are the measured chord-distributions
of the reconstruction for phase $j=1,2$ (at $M$ points).
The final reconstruction thus
has approximately the same chord functions as the experimental image
as well as sharing the low-order quantities $p$ and  $p^{(2)}(r)$. 

A limitation of this `model-based' technique is that one of the
of models tested must, for some choice of its morphological parameters,
be able to approximately reproduce the experimental microstructure.
For example, it would be unlikely that a model derived from
the iso-surfaces of a random field would be able to mimic
the highly structured morphology of randomly packed hard spheres.
In such a case we would expect that none of the model
chord functions would reproduce the experimental data.
At present there have not been a sufficient number of studies to
provide numerical criteria on $E\rho^{(j)}$ for acceptability.  
The general approach we have outlined is not restricted to
the models given in Sec.~\ref{statmod}. Ultimately it would
be useful to incorporate poly-disperse overlapping spheres
and other Boolean models such as those based on Poisson and Voronoi
polyhedra. The latter models have proved useful in the
analysis of mineralogical materials~\cite{KingBool} and
flow in porous filters~\cite{BourgeoisFilter}.
There may also prove to be more useful discriminants of
microstructure than the chord-functions.

To conclude this section we note that a
recent study~\cite{YeongRec1} has considered a
`model independent' scheme based on sequentially moving filled pixels
(representing the target phase) on a grid so that the
reconstruction reproduces statistical properties
of the original image. The method
differs from ours in that numerical estimates
of $p^{(2)}_{\rm{rec}}$ replace
$p^{(2)}_{\rm{mod}}$ in Eqn.~(\ref{Ep2}), and Eqns.~(\ref{Ep2}) and
(\ref{Epcd}) are coupled. The authors also employ the lineal
path distribution function $L^{(j)}(r)$ in Eqn.~(\ref{Epcd}) which
is related to the pore-chord functions
by $\rho^{(j)}(r)= \frac14 s_v d^2L^{(j)}(r)/dr^2$~\cite{TorqLu93}.

\section{Elastic properties}

\subsection{Theory}

The basic information
required for the evaluation of the effective moduli are the
volume fractions, and elastic moduli, of each phase;
$p$ (fraction of phase 1), $q=1-p$, $\kappa_i$ and $\mu_i$ ($i=1,2$).
A common approximation for the effective moduli is the self-consistent
method (SCM) of Hill~\cite{HillSCM} and
Budiansky~\cite{BudianskySCM}
which involves solving the equations
of elasticity for a spherical particle of phase 1 surrounded by a medium of
unknown effective moduli $\kappa_e$ $\mu_e$. The results are obtained
by solving
\begin{eqnarray}
\frac{p}{\kappa_e-\kappa_2}+\frac{q}{\kappa_e-\kappa_1}&=&
\frac3{3\kappa_e+4\mu_e} \\
\frac{p}{\mu_e-\mu_2}+\frac{q}{\mu_e-\mu_1}&=&
\frac{6(\kappa_e+2\mu_e)}{5\mu_e(3\kappa_e+4\mu_e)} 
\end{eqnarray}
for $\kappa_e$, $\mu_e$. In the case where one of the
phases is perfectly soft or rigid the results exhibit a percolation
threshold of $p=\frac12$. The formula is also symmetric
to phase interchange
[$\kappa_e(k_1,g_1,k_2,g_2,p)=\kappa_e(k_2,g_2,k_1,g_1,1-p)$ {\it etc.}].
These facts limit the applicability of the SCM, since most composites
have lower percolation thresholds and many are not
symmetric~\cite{Hashin83}. A more realistic formula is obtained
using a generalized SCM (GSCM)~\cite{ChristensenLoGSCM} for the case
of a particle of phase 1 surrounded by a spherical
shell of phase  (embedded in a medium of the effective moduli).
The result is complicated~\cite{Christensen90} and
not reproduced here. The GSCM
has zero percolation threshold, and is not symmetric under phase
interchange. For non-particulate media it is not clear which
phase should be associated with the inclusions  and which with
the matrix. Below we consider both cases.
Christensen~\cite{Christensen90} found that the GSCM
provided a better prediction of composite properties than other
common methods, so we shall not consider these here.
It should also be noted that the volume fraction is
the only microstructural information
included in the SCM and GSCM results. This means that these formulae
are insensitive to the distribution of each phase, or rather
that each formula has a ``built-in" microstructure, which may or
may not match the experimental one.

The difficulty in deriving general theoretical results for
predicting the elastic properties of random composites has provided the
impetus for the development of rigorous bounds~\cite{TorqRev91}.
For orientationly isotropic
materials the bounds take the general form~\cite{QuinRev98},
\begin{equation}
\left( \langle \kappa^{-1} \rangle
-\frac
{4pq(\kappa_2^{-1}-\kappa_1^{-1})^2 }
{4\langle\tilde{\kappa}^{-1}\rangle+3\Gamma}
 \right)^{-1}
\leq \kappa_e \leq
\langle\kappa\rangle-\frac{3pq(\kappa_2-\kappa_1)^2}
{3\langle\tilde{\kappa}\rangle+4\Lambda}
\label{genk}
\end{equation}
\begin{equation}
\label{genu}
\left( \langle \mu^{-1} \rangle
-\frac
{pq(\mu_2^{-1}-\mu_1^{-1})^2 }
{\langle\tilde{\mu}^{-1}\rangle+6\Xi}
 \right)^{-1}
\leq \mu_e \leq
\langle\mu\rangle-\frac{6pq(\mu_2-\mu_1)^2}
{6\langle\tilde{\mu}\rangle+\Theta}
\end{equation}
where for a variable $b$, $\langle b\rangle\equiv pb_1+qb_2$ and $\langle \tilde{b}
\rangle\equiv qb_1+pb_2$. The additional parameters $\Gamma$, $\Lambda$,
$\Xi$ and $\Theta$ depend on the level of
microstructural information available. 
If any of the moduli ($\kappa_i$ or $\mu_i$) are zero then
the lower bounds vanishes. Similarly if any of the moduli are
infinite the upper bound diverges.

If only the volume fractions of the composite are known
\begin{equation} \Gamma=\mu_1^{-1},\;\;\; \Lambda=\mu_2,\;\;\;
\Xi= \frac{\kappa_1+2\mu_1}{\mu_1(9\kappa_1+8\mu_1)} \;\;\; \& \;\;\;
\Theta=\frac{\mu_2(9\kappa_2+8\mu_2)}{\kappa_2+2\mu_2},
\end{equation}
and Eqns.~(\ref{genk})~\&~(\ref{genu}) are the bounds of Hashin
and Shtrikman~\cite{HSelas}
for the case $\mu_2\geq\mu_1$ and $\kappa_2\geq\kappa_1$.
The bounds only apply to well-ordered materials
$[(\kappa_2-\kappa_1)(\mu_2-\mu_1) \geq 0]$ and the inequality signs
in the bounds must be reversed if
$\mu_2<\mu_1$ and $\kappa_2<\kappa_1$.
If further information is available in the form of three-point
statistical correlations it is possible to
derive more restrictive bounds in terms of the
microstructure parameters~\cite{Milton81a},
\begin{eqnarray}
\label{zeta}
\zeta_1&=& 
\frac9{2pq}\int_0^\infty\!\!\frac{dr}{r} \int_0^\infty
\!\! \frac{ds}{s} \int_{-1}^1 du  P_2(u)
\left( p^{(3)}(r,s,t)-\frac{p^{(2)} (r) p^{(2)}(s)}{p} \right) \\
\label{eta}
\eta_1&=&\frac 5{21}\zeta_1+
\frac{150}{7pq}\int_0^\infty\!\!\frac{dr}{r}
\int_0^\infty\!\! \frac{ds}{s}
\int_{-1}^1 du  P_4(u) 
\left( p^{(3)}(r,s,t)-\frac{p^{(2)}(r) p^{(2)}(s)}{p}
\right).
\end{eqnarray}
where $t^2=r^2+s^2-2rs u$ and $P_2(u)=\frac12(3 u^2-1)$ and
$P_4(u)=\frac18(35u^4-30u^2+3)$ are Legendre polynomials.
In this case we have
$$
\Gamma=\langle\mu^{-1}\rangle_\zeta,\;\;\; \nonumber
\Lambda=\langle\mu\rangle_\zeta,\;\;\;
\Theta=\frac{3\langle \mu\rangle_\eta\langle 6\kappa+7\mu\rangle_\zeta-5\langle \mu\rangle_\zeta^2}
{\langle 2\kappa-\mu\rangle_\zeta+5\langle \mu\rangle_\eta} \;\;\; 
$$
\begin{equation}
\Xi=\frac{5\langle \mu^{-1}\rangle_\zeta\langle 6\kappa^{-1}-\mu^{-1}\rangle_\zeta+\langle \mu^{-1}\rangle_\eta
\langle 2\kappa^{-1}+21\mu^{-1}\rangle_\zeta}{\langle 128\kappa^{-1}+99\mu^{-1}\rangle_\zeta+45\langle \mu^{-1}\rangle_\eta}.
\label{BMMPpars}
\end{equation}
Here we have used the standard notation
$\langle b\rangle_\zeta\equiv \zeta_1b_1+\zeta_2b_2$ and
$\langle b\rangle_\eta\equiv \eta_1b_1+\eta_2b_2$ where $b_i$ is any
function of $\mu_i$ and $\kappa_i$, $\zeta_2=1-\zeta_1$ and $\eta_2=1-\eta_1$.
With the parameters given in Eqn.~(\ref{BMMPpars}), the bounds on $\kappa$
are due to Beran and Molyneux~\cite{Beran65b} while the
bounds on $\mu$ are those of Milton and Phan-Thien~\cite{Milton82b}.
The development of these bounds have been recently
reviewed~\cite{TorqRev91,QuinRev98}.
Below we consider the Young's modulus [$E=9\kappa\mu/(3\kappa+\mu)$]
and Poisson's ratio [$\nu=(3\kappa-2\mu)/(6\kappa+2\mu)$] of a
composite.
The bounds on $E$~\cite{Hashin83} and $\nu$~\cite{ZimmermanNu} are
\begin{equation}
\frac{9\kappa_l\mu_l}{3\kappa_l+\mu_l}\leq E_e \leq
\frac{9\kappa_u\mu_u}{3\kappa_u+\mu_u}
\;\;\;\&\;\;\;
\frac{3\kappa_l-2\mu_u}{6\kappa_l+2\mu_u}\leq \nu_e \leq
\frac{3\kappa_u-2\mu_l}{6\kappa_u+2\mu_l},
\label{ESbounds}
\end{equation}
where the subscripts refer to the upper and lower bound of
Eqns.~(\ref{genk})~\&~(\ref{genu}).
Depending
on the level of microstructural information employed
to find the bounds on $\kappa$ and $\mu$ we refer to Eqn.~(\ref{ESbounds})
as the Hashin and Shtrikman (HS) or Beran, Molyneux, Milton and
Phan--Thien (BMMP) bounds respectively.
The microstructure parameters $\zeta$ and $\eta$ have been
evaluated for hard and overlapping spheres~\cite{TorqRev91},
level cut Gaussian random-field models~\cite{Roberts95a}
and can be evaluated for other Boolean models~\cite{Jeulin97a}.


\subsection{Computation}

A microstructure made up of a digital image is already naturally discretized
and so lends itself to numerical computation of many quantities.  For computing
elastic moduli, there are two methods available:  a finite element
method~\cite{Garboczi95a}, and a finite difference method~\cite{Adlerelas1}.
The finite element method uses a variational formulation of the linear elastic
equations, and finds the solution by minimizing the elastic energy via a fast
conjugate gradient method.  The finite difference method formulates the linear
elastic equations directly in a finite difference approach, and solves the
resulting set of linear equations with a similar conjugate gradient method.

For a porous material, with one solid phase and one pore phase, either method
can be used, as the zero normal force boundary condition at a solid-pore
boundary is easy to handle in either method. When there are solid-solid
boundaries between two different phases, the boundary conditions become 
continuity of displacement and continuity of normal force. This is harder to
implement in the finite difference method, while it is just as easy to do
as in the solid-pore case for the finite element method. We have used the 
finite element method exclusively in this paper. 

The finite element method is one that has been especially adapted for digital
images.  It is for linear elasticity only. Each pixel, in 3-D, is taken 
to be a tri-linear finite element~\cite{Cook}.
There can be any number of phases, whether isotropic or anisotropic,
as long as each phase can be adequately depicted within the resolution of the 
digital image used, and can be described with a single elastic moduli tensor.
Thermal strains can also be easily handled ~\cite{Eds_manual}.  
The digital image is assumed to
have periodic boundary conditions.  A strain is applied, with the average
stress or the average elastic energy giving the effective elastic
moduli~\cite{TorqRev91,Hashin83}. Details of the theory and programs used are
reported in the papers of Garboczi \& Day~\cite{Garboczi95a}
and Garboczi~\cite{Eds_manual}. The actual programs are
available at {\em http://ciks.cbt.nist.gov/garboczi/}, Chapter 2. 

\subsection{Test case: Overlapping sphere model}

\begin{figure}[bt!]
\centering\epsfig{figure=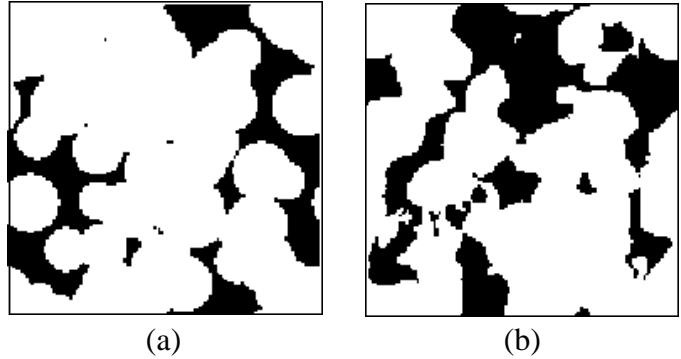,width=9.0cm}
\caption{Cross-sections of (a) the overlapping sphere model, and (b) the
best reconstruction (model I$_{10}$ $c$=0). The volume fraction is $p$=20\%
and the images are 96$\times$96 pixels.
\label{recios}} 
\end{figure}
\begin{figure}[bt!]
\centering\epsfig{figure=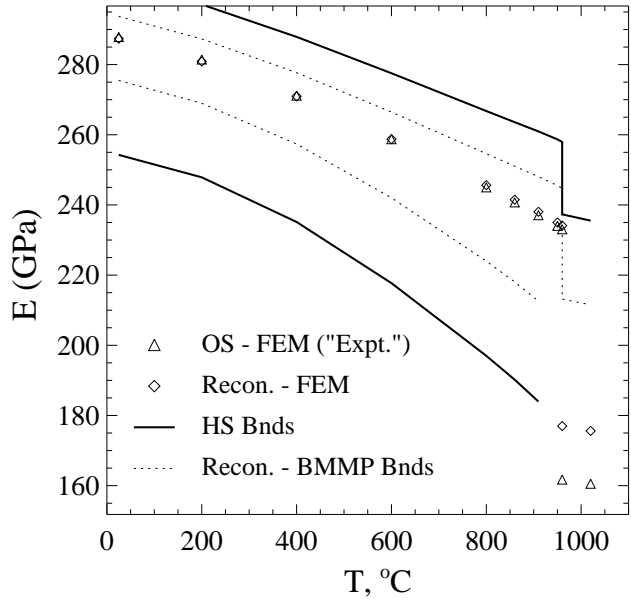,width=8.5cm}
\caption{Young's modulus of the overlapping sphere model (OS) compared
with data obtained from the best reconstruction (Recon.) [model I$_{10}$ ($c$=0)] (Finite
element method). The Beran, Molyneux, Milton and Phan--Thien (BMMP) bounds
are seen to be more restrictive than
the Hashin and Shtrikman (HS) bounds for both models.
\label{elasios}}
\end{figure}

Before going ahead and reconstructing the W--Ag composite we first tested
the procedure for the over-lapping sphere model for which
the statistical properties can be analytically evaluated~\cite{TorqStel83}.
A 3-D realization of this model is also easy to generate,
so that the reconstruction can be carefully tested.
The overlapping sphere
model [Fig~\ref{recios}(a)] has previously been reconstructed
using the models derived from the level cut
GRF's~\cite{RobertsRec}. Model I$_{10}$ ($c$=0) was found to
be the best reconstruction [Fig~\ref{recios}(b)].
To gauge the accuracy of the
procedure for the elastic properties of the tungsten-silver composite
(see the following section)
we have computed the elastic moduli of the overlapping
sphere model and its reconstruction at volume fraction $p=20\%$ (of the phase
outside
the spheres).
The moduli of each phase at each temperature is set to
the corresponding value for silver and tungsten.
The results, shown in Fig.~\ref{elasios}, indicate that the
procedure performs very well. When the silver has non-zero elastic moduli,
for temperatures below the melting point of silver, the
reconstruction provides an extremely good prediction
of $E$ (error $<$1\%). At 
temperatures above the melting point of silver, the silver is taken
to have zero shear and bulk moduli, and the
reconstructed model is 9\% stiffer than overlapping spheres.
Since the moduli depend most strongly
on microstructure at high contrast (contrast = the ratio of
the Young's moduli between the two phases) the latter
error is likely to be more indicative of the ability of
the model to reproduce the microstructure of overlapping spheres.
Nevertheless the reconstruction provides a reasonable model. Similar agreement
was seen for the Poisson ratios (not shown).

An advantage of the method is that
the BMMP bounds [Eqn.~(\ref{ESbounds})] can be evaluated
(since $p^{(3)}_{\rm{rec}}$ can be computed). 
The microstructure parameters
for the reconstruction [model I$_{10}$ ($c$=0)] were estimated as
$\zeta_1=0.43$, $\eta_1=0.35$~\cite{RobertsRec}.
The bounds (see Fig.~\ref{elasios}) are significantly more
restrictive than those of Hashin and Shtrikman, and
are seen to bound the finite element moduli of both
the overlapping sphere and reconstructed models.
Above the melting point the lower bounds are identically zero,
as we have taken the elastic properties of silver to be zero
at this point. Actually, above the melting point of silver, it would
be more reasonable to suppose that the silver has a non-zero
bulk modulus, with a zero shear modulus. This is only important
when comparing with experimental results, however,
and not in this model-model comparison.

\section{Application to a tungsten-silver composite}
\label{WAg}

To quantitatively test the reconstruction method, experimental 
data need to be available giving a picture of the material, properties for
each individual phase, and overall composite properties.
A well characterized system, suitable to test the reconstruction
procedures, is provided by the tungsten-silver
(W--Ag) composite of
Umekawa {\em et al.}~\cite{Umekawa65}.
This composite was 
produced by infiltrating a porous tungsten solid with
molten silver (volume fraction of silver $p$=20\%). The Young's
modulus of the composite was measured at
a range of temperatures above and below the melting
point of silver (960$^o$C). 
The elastic moduli of
each phase were obtained
by measuring the moduli of pure samples of tungsten
and silver at each temperature.
This data cannot be used directly because both phases of the composite
actually contained tiny spherical pores. These will reduce the 
Young's moduli and Poisson's ratio of each phase.
This effect can be accounted for by applying 
well known results for dilute spherical inclusions~\cite{Christensen90}.
For porous materials (porosity $\phi\ll1$) the formulae can be rewritten
as
\begin{eqnarray}
E_\phi&=&E_m-\phi E_m\left( \frac{9-4\nu_m -5 \nu_m^2}{7-5\nu_m} \right) \\
\nu_\phi&=&\nu_m-\frac32 \phi \left(\frac{(5 \nu_m-1)(1-\nu_m^2)}{7-5\nu_m}
\right)
\end{eqnarray}
where $E_m$, $\nu_m$ denote matrix properties and $E_\phi$, $\nu_\phi$ are
the porosity modified values.
The tungsten matrix had an internal porosity of 1\% while
the silver phase had a porosity of 10\% at room temperature,
decreasing linearly to 5\% at the melting point.
Table~\ref{table4} shows the phase moduli used at different temperatures.
 
\begin{table}[bt!]
\caption{The moduli of the silver and tungsten phases, as a function of 
temperature, after being corrected for the internal porosity of each phase.
\label{table4}}
\begin{center}
\begin{tabular}{ccccc}       
\hline
\multicolumn{1}{c}{} &
\multicolumn{2}{c}{Silver} &
\multicolumn{2}{c}{Tungsten} 
\\
\multicolumn{1}{c}{Temp($^o$C)} &
\multicolumn{1}{c}{E(GPa)} &
\multicolumn{1}{c}{$\nu$} &
\multicolumn{1}{c}{E(GPa)} &
\multicolumn{1}{c}{$\nu$} \\
\hline

25    & 71 & 0.36 & 400 & 0.28  \\
200   & 69 & 0.36 & 392 & 0.28  \\
400   & 63 & 0.36 & 383 & 0.28  \\
600   & 54 & 0.36 & 373 & 0.28  \\
800   & 45 & 0.37 & 363 & 0.28  \\
860   & 42 & 0.37 & 361 & 0.28  \\
910   & 39 & 0.37 & 359 & 0.28  \\
950   & 37 & 0.37 & 357 & 0.28  \\
960   & 37 & 0.37 & 356 & 0.28  \\
960   &  0 & 0.50 & 356 & 0.28  \\
1020  &  0 & 0.50 & 354 & 0.28  \\
\hline
\end{tabular} 
\end{center}
\end{table}

\begin{figure}[bt!]
\centering\epsfig{figure=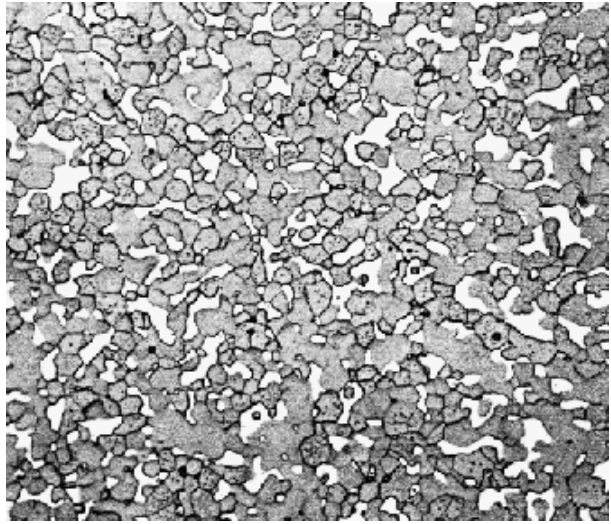,width=8cm}
\caption{The original scanned image from
Umekawa {\em et al.}~\protect\cite{Umekawa65}
(236$\times$204$\mu$m at 759$\times$657 pixels).
The dark phase corresponds to tungsten and the lighter to silver.
\label{digtag}}
\end{figure}

\begin{figure}[bt!]
\centering\epsfig{figure=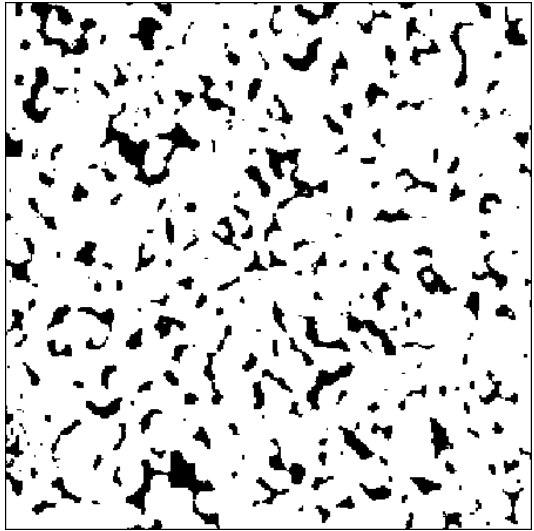,width=7.0cm}
\caption{The cropped (640$\times$640 pixels) binary image
obtained from the scanned image. This is the sample used
to calculate the statistics of the composite (side length
198.7$\mu$m). The black phase corresponds to silver.
\label{bwtag}}
\end{figure}

\begin{figure}[hbt!]
\centering\epsfig{figure=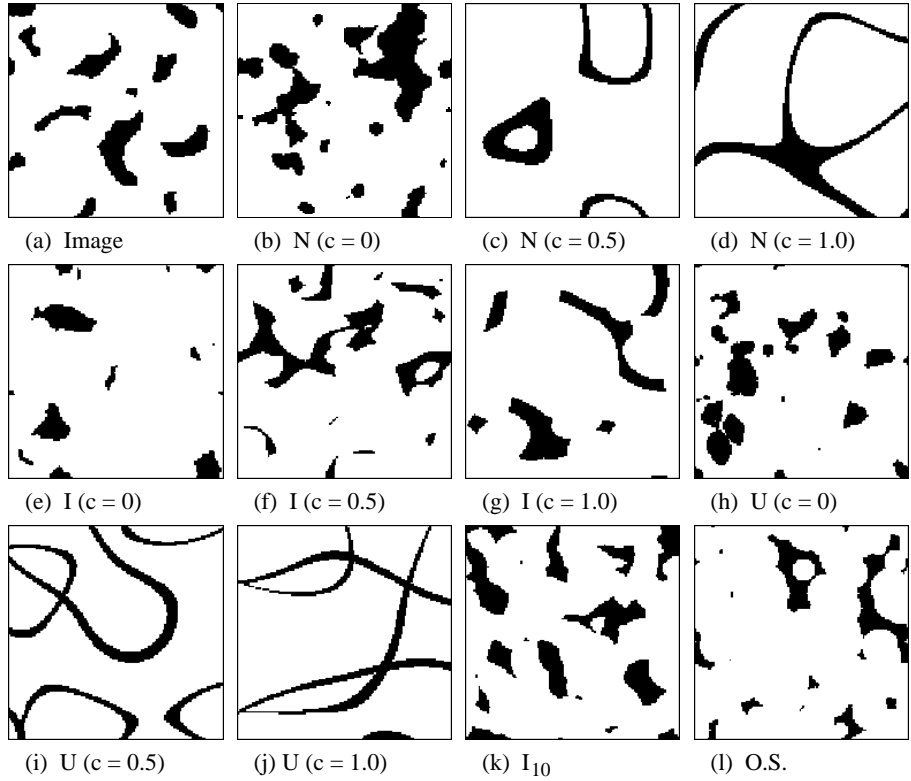,width=12.0cm}
\caption{Cross-sections of a portion of the original image (a)
and the eleven trial reconstructions (b-l) at $p$=13.5\%. The length-scale
parameters of the trials are chosen to match $p^{(2)}_{\rm{expt}}(r)$.
The chord-distribution's (Table~\protect\ref{tab_recon}) indicate that model
N ($c$=0) [shown in (b)] provides the best reconstruction.
Each image has side length
$39.7\mu$m.
\label{all_rec}}
\end{figure}

\begin{figure}[hbt!]
\centering\epsfig{figure=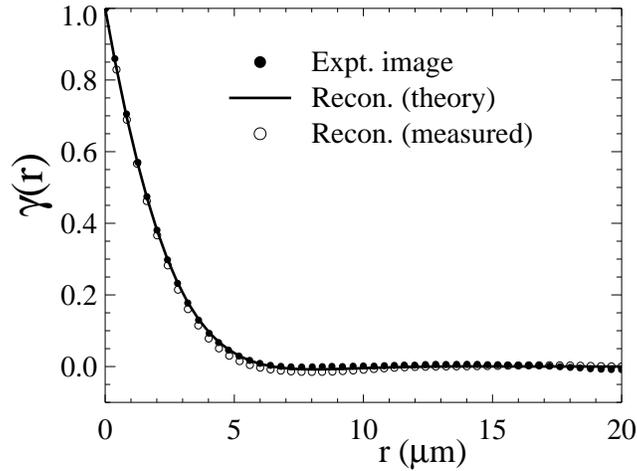,width=8.5cm}
\caption{The auto-correlation function of the
experimental image is well reproduced by the best reconstruction
[model N ($c$=0)].
\label{cf_p2}}
\end{figure}

\begin{figure}[hbt!]
\centering\epsfig{figure=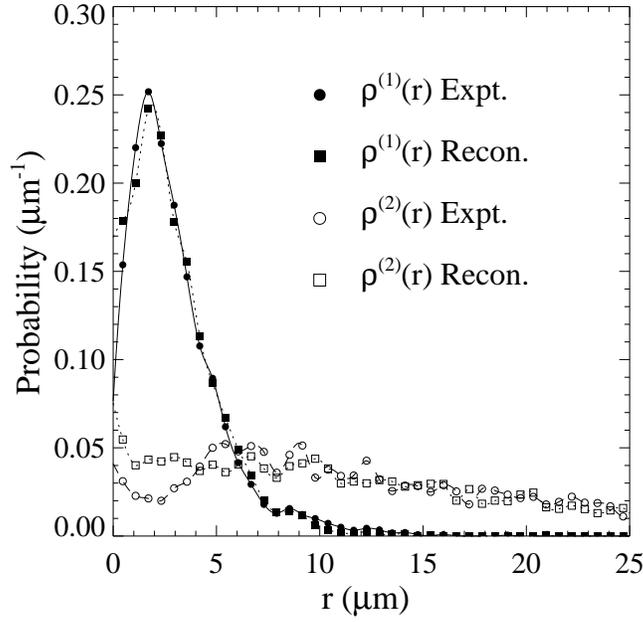,width=8.5cm}
\caption{The good agreement between the model and experimental
chord distributions [silver:$\rho^{(1)}(r)$; tungsten:$\rho^{(2)}(r)$]
indicates that model N ($c$=0) provides the best reconstruction.
\label{cf_po}}
\end{figure}

\begin{figure}[hbt!]
\centering\epsfig{figure=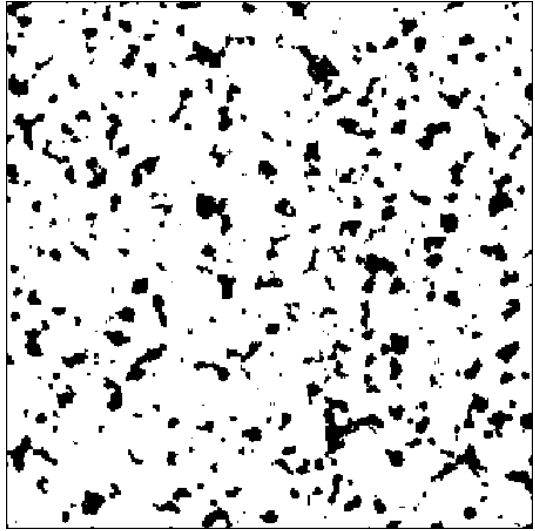,width=7.0cm}
\caption{The best reconstruction [model N ($c$=0)]. The region
shown is 198.7$\times$198.7$\mu$m (cf.\ Fig.~\protect\ref{bwtag}).
\label{2D_NC0}}
\end{figure}

\begin{figure}[hbt!]
\centering\epsfig{figure=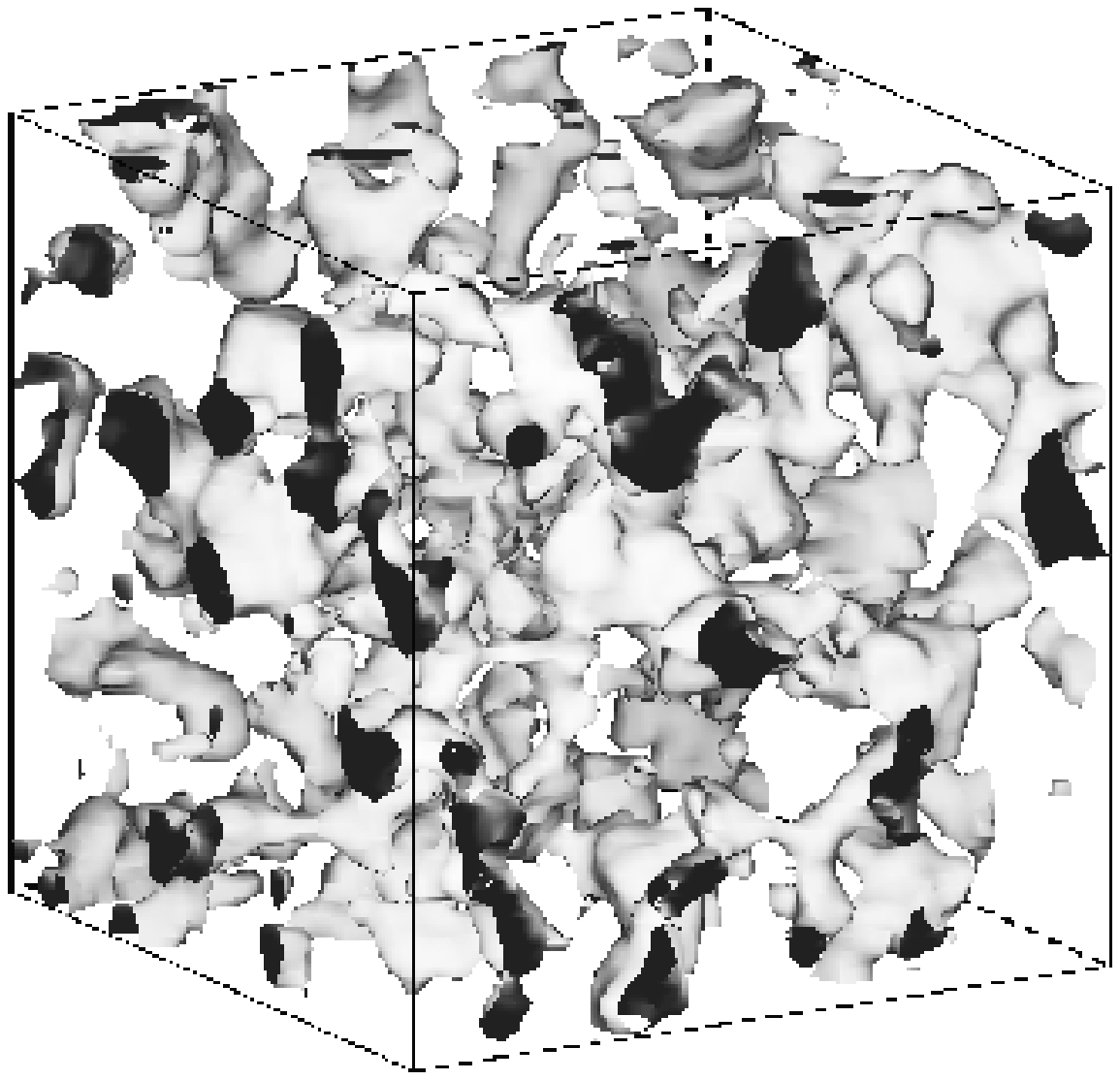,width=8.0cm}
\caption{The silver phase (shown as solid) of the
best reconstruction [model N ($c$=0)]. The side length
is 39.7$\mu$m. 
\label{3D_NC0}}
\end{figure}

To reconstruct the W-Ag composite we digitize a photograph
of the sample (Fig.~\ref{digtag}). All points below
a selected threshold grey-level are set to black (the silver phase)
while the remainder is set to white. The image was blurred and
re-thresholded to remove the pores of the tungsten
matrix (which appeared as silver). This had little effect
on $p_{\rm{expt}}$ and $p^{(2)}_{\rm{expt}}$, but
a significant effect on the measured silver chord distribution.
The resulting image is shown in Fig.~\ref{bwtag}.
The image actually has a silver content of only 13.5\%, significantly
lower than the nominal value of 20\%.
The statistical properties of the image are compared with
those of 11 trial reconstructions in Table~\ref{tab_recon}, while
2-D slices of the models themselves are shown, for purposes of 
illustration, in Fig.~\ref{all_rec}.  Several
of the models were unable to reproduce $p^{(2)}_{\rm{expt}}$
and were considered no further. A comparison of the chord distributions
indicated that model N ($c$=0) provided the best reconstruction.
The auto-correlation function 
of this model is compared with experimental data in Fig.~\ref{cf_p2}.
The experimental and model chord-distributions are shown in Fig.~\ref{cf_po}.
The silver distribution $\rho^{(1)}_{\rm{expt}}$ is well reproduced
by the model at all lengths shown, while $\rho^{(2)}_{\rm{mod}}$ performs
less well at small chord lengths.
Two and three dimensional images of the model
are shown in Fig.~\ref{2D_NC0} (shown at the same scale
as Fig.~\ref{bwtag})  and in Fig.~\ref{3D_NC0}.
\begin{table}[hbt!]
\caption{A comparison of the statistical properties of 11
reconstructions with those of the experimental composite;
$p$=13.5\% and $s_v$=0.17$\mu$m$^{-1}$
(obtained from Fig.~\protect\ref{bwtag}).
Many of the models are able to reproduce the low order
statistical properties of the composite ($Ep^{(2)}<0.1$). This shows
that $p^{(2)}(r)$ does not uniquely specify composite microstructure.
\label{tab_recon}}
\begin{center}
\setlength{\extrarowheight}{2pt}
\begin{tabular}{llccccccc}       
\hline
Mod.\ &$c$& $r_c$ & $\xi$ & $d$ & $s_v$ & $Ep^{(2)}$ &
$E\rho^{(1)}$&$E\rho^{(2)}$ \\
\hline
N&0        &2.16&2.15&13.0&0.19&0.05&0.03&0.28 \\
N&$\frac12$&28.1&28.2&22.0&0.18&0.11&    &      \\
N&1        &\iny&\iny&25.6&0.18&0.10&    &      \\
I&0        &2.88&2.89&12.5&0.20&0.05&0.28&0.88  \\
I&$\frac12$&13.4&13.5&15.9&0.18&0.06&0.32&0.53  \\
I&1        &32.7&32.1&17.4&0.17&0.05&0.11&0.38  \\
U&0        &2.69&2.68&13.1&0.18&0.05&0.08&0.30  \\
U&$\frac12$&60.2&144 &35.5&0.20&0.15&    &      \\
U&1        &\iny&\iny&43.0&0.20&0.15&    &      \\
I$_{10}$& 0   &4.75&4.76&12.6&0.21&0.06&0.33&0.58  \\
\multicolumn{2}{c}{OS} &
\multicolumn{3}{c}{$r_0$=3.75} &
\multicolumn{1}{c}{0.21} &
\multicolumn{1}{c}{0.25} &
\multicolumn{1}{c}{0.49} &
\multicolumn{1}{c}{0.45} \\
\hline
\end{tabular} 
\end{center}
\end{table}

\begin{figure}[hbt!]
\centering\epsfig{figure=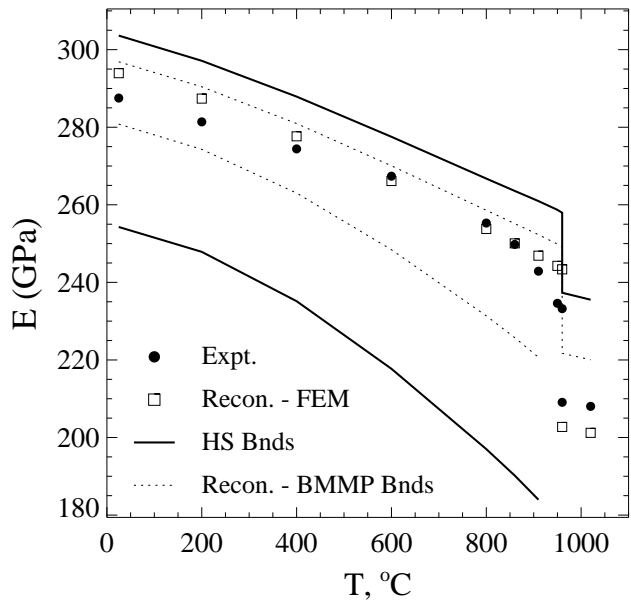,width=8.5cm}
\caption{The experimental Young's moduli compared with
FEM data obtained from the best reconstruction [Fig.~\protect\ref{3D_NC0}].
The Beran, Molyneux, Milton and Phan--Thien (BMMP) bounds
for the reconstruction
and Hashin and Shtrikman bounds are also shown.
\label{cf_young}}
\end{figure}

For the
purposes of computing the elastic properties of the
model we maintain the length scale parameters ($\xi$, $r_c$ and $d$)
and alter the level cut parameters ($\alpha$ and $\beta$) of
the model such that $p_{\rm{rec}}$=20\%
(in accord with the experimental composite).
The Young's modulus, computed using the finite element method,
is compared with the experimental
data of Umekawa {\em et al.} in Fig.~\ref{cf_young}.
For the temperature region below the melting point of silver
the maximum error is 4\%, a very good result. Above the melting
point of silver, when the silver phase is taken to have a zero bulk and
shear modulus, the error is only 3\%.  The agreement may actually be
better than that, however.  Since the elastic measurements were
dynamic measurements, the liquid silver can be considered as being
trapped on the time scale of the experimental measurement, before
any significant flow could take place, and so could still contribute to
the effective moduli via 
its non-zero liquid bulk modulus.  Just before melting, the silver
had a bulk modulus of about 35GPa.  If we take the bulk
modulus to be somewhat lower, in analogy to the ice-water difference around
0$^o$C, then a bulk modulus of 23.1GPa causes
the N ($c$=0) model to agree perfectly
with experiment at temperature points above the melting point of silver.

The bounds are also shown in Fig.~\ref{cf_young}.
For model N ($c$=0) the microstructure parameters
are $\zeta_1=0.31$ and $\eta_1=0.27$.
The results bound the experimental
data and provide a reasonable prediction of the Young's modulus
below the melting point of silver. 
Note that even if the silver phase is given a non-zero bulk modulus past
its melting point, the zero shear modulus causes the lower bounds for shear
modulus and therefore Young's modulus to be identically zero.
Unfortunately, there was no reported Poisson's ratio results for the
composite, so we cannot compare to the model results for this quantity.

\section{Analysis of analytical effective elastic moduli predictions}
\label{EMT}

\begin{figure}[htb!]

\begin{center}
\begin{minipage}[bt!]{8.5cm}

\noindent
(a)
\vspace{-10mm}

\begin{center}
\epsfig{figure=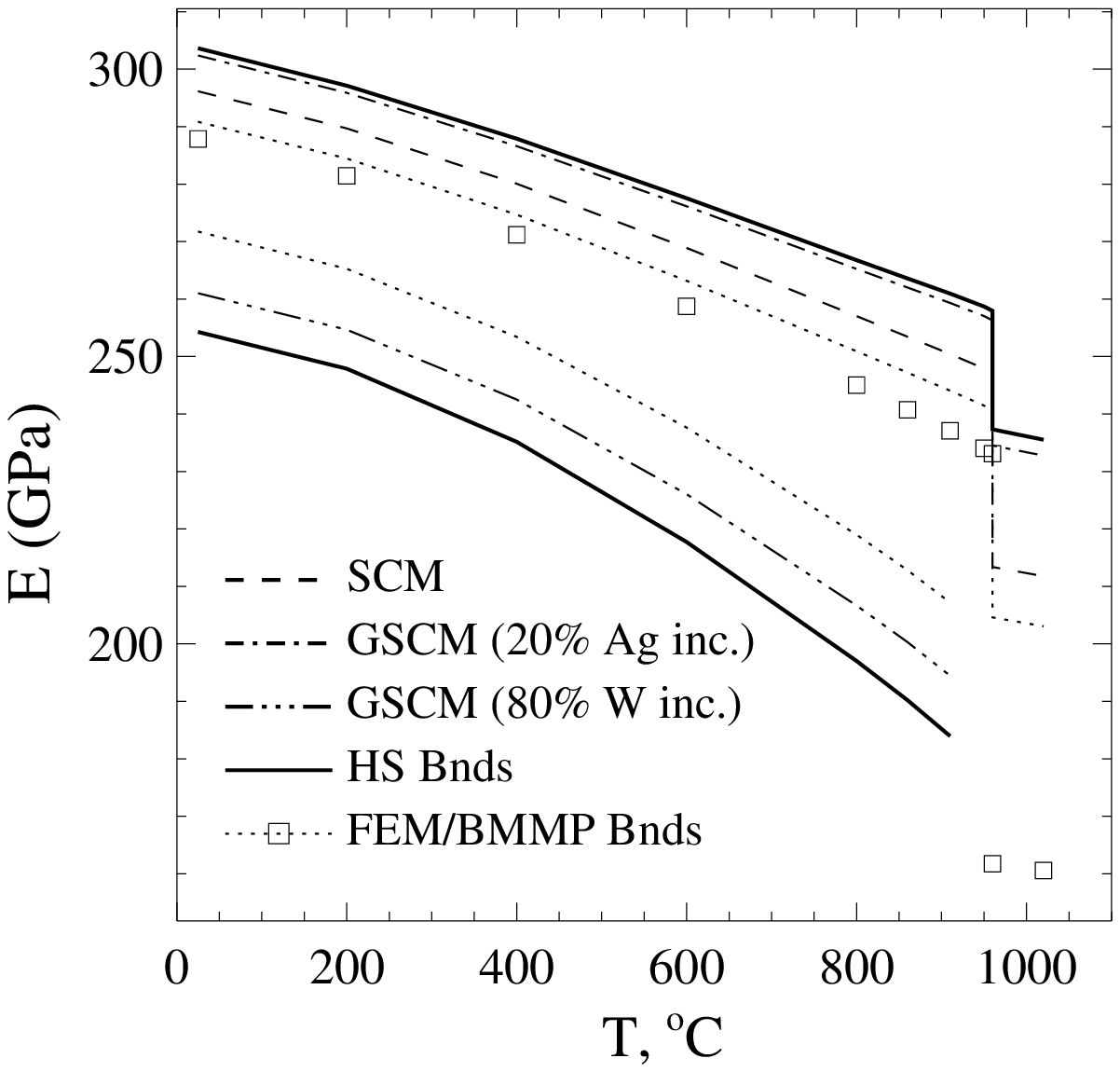,width=\linewidth}
\end{center}

\vspace{-3mm}

\noindent
(b)
\vspace{-10mm}

\begin{center}
\epsfig{figure=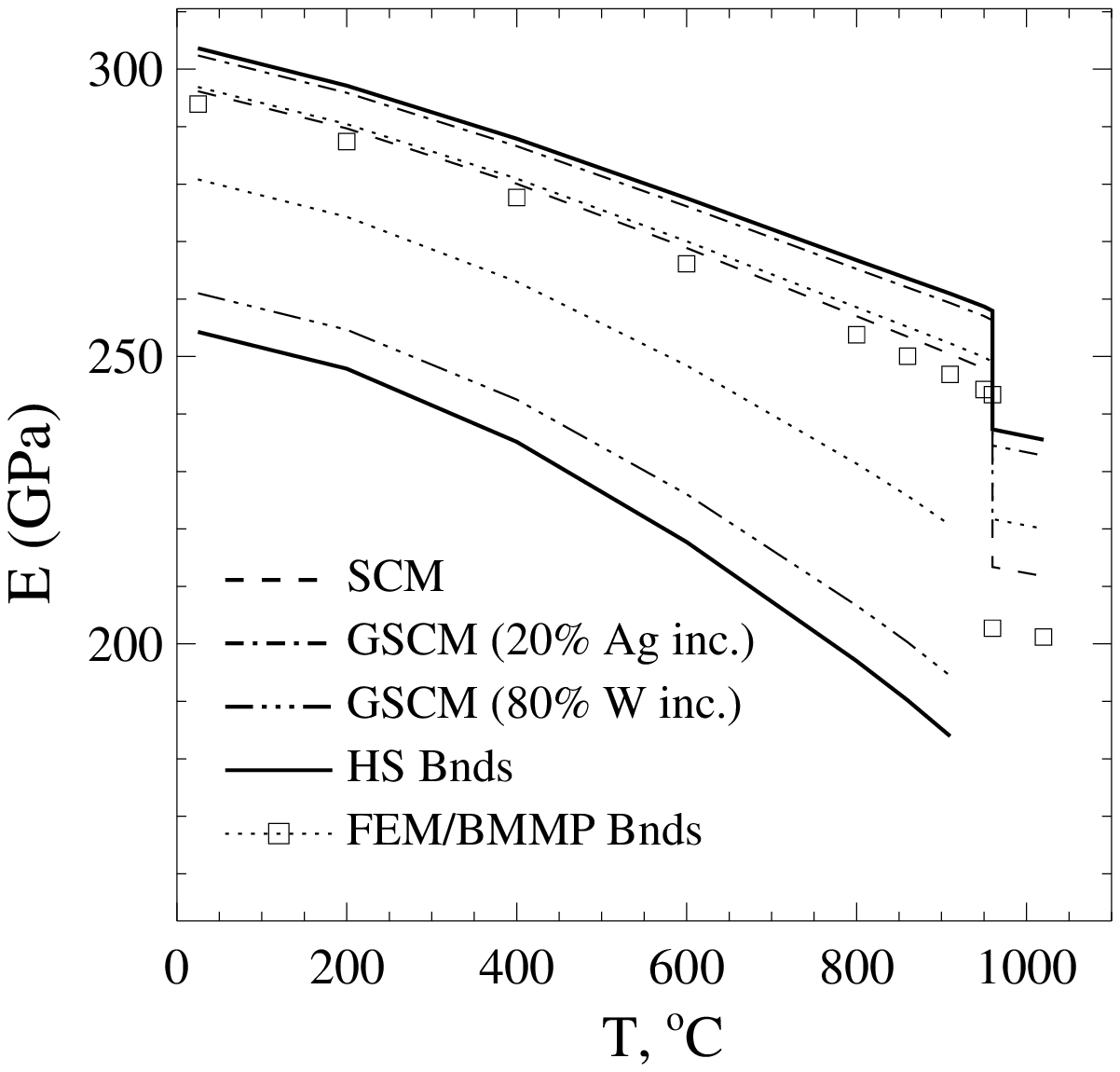,width=\linewidth}
\end{center}

\end{minipage}
\end{center}

\vspace{1mm}

\caption{Comparison of theory (predictions and bounds) with finite element
(FEM) calculations for the Young's modulus of (a) the overlapping
sphere model and (b) the single-cut GRF model [N($c$=0), see Fig.~\protect\ref{3D_NC0}].
The standard (SCM)
and generalized (GSCM) self-consistent methods are shown, as are the
Hashin and Shtrikman (HS) and Beran, Molyneux, Milton, and Phan--Thien (BMMP) bounds.
\label{cf_thy}} 
\end{figure}

We now compare the different analytical predictions of effective moduli
with the finite element
data. We have chosen to study the two most common models
and only consider the moduli appropriate for the W-Ag composite
studied above.
The results are shown in Fig.~\ref{cf_thy}(a) for overlapping spheres
and in Fig.~\ref{cf_thy}(b) for the single level-cut GRF
or excursion set of Quiblier [model N($c$=0)]. 
The self-consistent method provides a very good estimate
of $E_e$ for model N($c$=0), but not for overlapping
spheres. This might be expected because the N($c$=0) model
is symmetric with respect to phase-interchange (like the SCM) while
the overlapping sphere model is not.
As stated above the application of the
generalized SCM is difficult because it
is not obvious which phase should be chosen as the `inclusion' phase.
For the overlapping sphere model the tungsten phase is comprised of
spheres (at 80\% volume fraction), so is the more likely 
choice for the inclusion phase. Nevertheless,
we report both estimates (80\% W inclusions and 20\% Ag matrix
or 20\% Ag inclusions and 80\% W matrix) for both models. For either choice,
the GSCM fails to provide an accurate estimate.
Indeed, above the melting point of silver the GSCM vanishes for
the case 20\% Ag matrix case since the
matrix phase is now completely soft.
For the overlapping
sphere model the
Beran, Molyneux, Milton and Phan--Thien
bounds are calculated using
the microstructure parameters
$\zeta_1=0.52$, $\eta_1=0.42$~\cite{TorqStel83,Berryman85}.
Below the melting point
of silver (where the contrast between the phases is moderate)
the upper bounds provide a very good estimate of the
effective moduli.

A brief discussion of the effect of elastic contrast is necessary here.
We have already noted that the analytical predictions of effective moduli
do not explicitly depend on microstructure, but have a ``built-in"
microstructure. The elastic contrast, the ratio between the phase moduli,
will determine how sensitive the effective moduli actually are to
microstructure.  For example, in the case of a two-phase composite
having equal shear moduli but different bulk moduli, there is a simple 
exact formula for the effective bulk modulus which is totally insensitive
to microstructure~\cite{Hillequal}. In the case of small contrast, 
the effective 
moduli can be expressed exactly as a power series in the moduli differences
~\cite{Torqpower}.
Up to second order in this difference, at any volume fraction,
the coefficients of the power series
are not dependent on anything but the volume fractions and the individual
phase properties. Therefore at small contrast, 
analytical predictions of effective moduli that explicitly depend only on
volume fractions and phase moduli should all work well.

\clearpage

\section{Conclusion}

Throughout this paper, we have treated the finite element computation method 
as being perfectly accurate, so that comparisons of elastic results to 
experiment were solely a test of how well the reconstructed microstructure 
compared to the real microstructure. This is not exactly true, since there
are numerical errors in the finite element
method ~\cite{Garboczi95a,Eds_manual}.  
These are small, however,
and are generally of about the same size or less than the differences seen 
between model computations and experimental data for the elastic moduli. 
There are also statistical sampling errors associated with the
finite size ($\approx$40$\mu$m) of the models we employ to estimate the
elastic properties. Since this is much greater than the correlation
length of the samples ($\approx$5$\mu$m - see Fig.~\ref{cf_p2}) we
again assume these errors to be small.
Therefore, the good agreement between model prediction and experimental data
seen in this paper is good evidence that the model considered is indeed 
capturing the main aspects of the experimental microstructure.

We have compared various theoretical results to finite element
computations of the effective Young's modulus $E_e$ for non-particulate
media: a W-Ag composite and two model media (overlapping spheres and a
single-cut Gaussian random field). 
The generalized self-consistent method
(derived for particulate composites) did not provide a good
estimate of $E_e$ for the bi-continuous materials considered here.
The standard self-consistent method provided a good estimate for the single-cut
GRF and W-Ag composite. Since the method predicts zero moduli
for porosity above 50\% but the solid phase of the
single-cut GRF remains connected up to porosities of around
90\%~\cite{Roberts95a} such agreement cannot be general.
Upper bounds, calculated using three-point statistical correlation
functions, provided a good prediction at low contrast
($E_1$/$E_2\approx$ 6) for each composite. When one of the phases
was completely soft the bounds lost predictive value.
Therefore, for general composites, it is important to
employ numerical computations of the effective moduli.
For accurate numerical prediction of composite properties it
is important that a realistic model be used. Model-based
statistical reconstruction, based on the Joshi-Quiblier-Adler
approach, appears to be a viable route
for microstructural simulation. However, it is important that the models
underlying the procedure be capable of mimicking the composite microstructure.
We have shown how several different models can be employed to find
a useful reconstruction.

\begin{center} {\sc Acknowledgments} \end{center}

\noindent
A.R.\ thanks the Australian-American Educational Foundation (Fulbright
Commission) for financial support and the Department of Civil
Engineering and Operations Research at Princeton University where
this work was completed.  We also thank the Partnership for
High-Performance Concrete program of the National Institute
of Standards and Technology for partial support of this work.


\end{document}